\documentclass[%
 reprint,
 amsmath,amssymb,
 aps,
]{revtex4-2}

\usepackage{graphicx}
\usepackage{dcolumn}
\usepackage{bm}
\usepackage{hyperref}
\usepackage{xcolor}
\usepackage{float}

\hypersetup{
    colorlinks=true,
    citecolor=blue,
    linkcolor=black,
    urlcolor=blue,
}

\begin{document}

\preprint{APS/123-QED}

\title{Emergence of isotropy in rotating turbulence of Bose-Einstein condensates}

\author{Yuto Sano}
\affiliation{
 Department of Physics, Osaka City University, 3-3-138 Sugimoto, Sumiyoshi-ku, Osaka 558-8585, Japan}

\author{Makoto Tsubota}
\affiliation{%
 Department of Physics, Nambu Yoichiro Institute of Theoretical and Experimental Physics (NITEP), Osaka Metropolitan University, 3-3-138 Sugimoto, Sumiyoshi-ku, Osaka 558-8585, Japan
}%

\date{\today}
\begin{abstract}
We present a study on the development of rotating turbulence in Bose-Einstein condensates with a dissipative Gross-Pitaevskii model. Turbulence is generated by driving the lattice of quantized vortices in a harmonic potential with a random forcing potential. As the turbulence progressed, the initial alignment of vortices underwent slight disruptions, thereby increasing the high-wavenumber components of the kinetic energy. In the turbulent state, the distribution of incompressible kinetic energy exhibits milder anisotropy than that in the initial lattice state and demonstrates a scaling behavior of $k_z^{-2.5}$ in the direction parallel to the rotation axis. In contrast, the compressible kinetic energy exhibits an isotropic scaling behavior at high wavenumbers. 
\end{abstract}

\maketitle

\section{\label{sec:level1} Introduction}

Rotating turbulent flow is a common occurrence in nature, and the presence of rotation provides turbulence a characteristic behavior. An example of such behavior is the anisotropic energy transfer influenced by the Coriolis force, which results in an asymmetric energy transfer between the directions parallel and perpendicular to the axis of rotation; it has been thoroughly investigated in classical turbulence \cite{Greenspan1968,Waleffe1993,Zeman1994,Cambon1997,Galtier2003,Staplehurst2008,Moisy2011,Lamriben2011,Sharma2019}. The influence of rotation on turbulence is quantified by the Rossby number ${\rm Ro}=U/(2\Omega l)$, where $U$ represents the velocity scale corresponding to the integral length scale $l$ of the turbulence, and $\Omega$ denotes the angular frequency. When ${\rm Ro} < 1$, the kinetic energy distribution of isotropic turbulence is more concentrated perpendicularly rather than parallelly, resulting in a coherent redistribution of vorticity along the rotation axis. The anisotropic characteristics due to the Coriolis force are evident at wavenumbers smaller than the Zeman wavenumber $k_{\Omega}=(\Omega^3/\epsilon)^{1/2}$ ($\epsilon$ is the kinetic energy dissipation), where inertial waves are dominant and the kinetic energy spectrum exhibits a $k^{-2}$ scaling steeper than a Kolmogorov spectrum $k^{-5/3}$ \cite{Zhou1995,Canuto1997,Yeung1998,Baroud2002}. On the other hand, above wavenumbers than $k_{\Omega}$, the effect of rotation diminishes, and the inertial subrange of the isotropic Kolmogorov turbulence can be restored.

Rotation plays an important role in the dynamics of the atomic Bose-Einstein condensate (BEC), resulting in the formation of quantized vortices with quantized circulation $\kappa=h/m$ (where $m$ represents the mass of atoms) \cite{Pethick2001}. In the rotating frame with $\Omega$, the Hamiltonian can be written as $H'=H-\Omega L$ where $H$ is the Hamiltonian in the non-rotating frame, and $L$ is the total angular momentum along the rotation axis. Above the critical angular frequency, the vortex states exhibit greater stability than a state devoid of vortices. In such instances, some quantized vortices can penetrate the condensate and form a vortex lattice along the rotation axis. Such dynamics in rotating BECs have been investigated through several experiments \cite{ madison2000,Abo2001,haljan2001,hodby2001,Fetter2009} and numerical calculations \cite{tsubota2002,penckwitt2002,kasamatsu2003,kasamatsu2005} using the Gross-Pitaevskii (GP) model. The GP model represents a non-perturbative mean-field equation for a classical field that captures the dynamics of weakly interacting BECs quantitatively.

The BEC provides a highly controllable platform for investigating turbulence. The unique characteristics of this system offer valuable opportunities to explore diverse forms of turbulence, including vortex turbulence \cite{nore1997,nore1997fluids,Kobayashi2005,Kobayashi2005jps,Kobayashi2007,Bagnato2009,tepez2009}, turbulence in two-component BEC \cite{Takeuchi2010,Kobyakov2014,Mithun2021}, and wave turbulence \cite{Lvov2003,Zakharov2005,Nazarenko2006,Nazarenko2007,Proment2009,Proment2012,Fujimoto2015,Nir2016,Nir2019,Shukla2022,Bagnato2022,Galka2022,Sano2022,Zhu2023,Dogra2023}. Recent studies observed the emergence of statistical isotropy in the momentum distribution during the development of quantum turbulence despite the anisotropic initial conditions and energy injection \cite{Galka2022,Sano2022}. The emergent isotropy is one of the similarities between classical and quantum turbulence and has been abundantly studied in classical turbulence \cite{lumley1977,yeung1991,frisch1995,choi2001,davidson}. Furthermore, it forms the backbone of various turbulence theories, as exemplified by Kolmogorov's theory, and is a universal characteristic of turbulence independent of classical and quantum considerations.

The investigation of rotating quantum turbulence (RQT) remains relatively limited \cite{Swanson1983,Tsubota2003,Tsubota2004,Hossain2022,Estrada2022,Estrada2022_2,Makinen2023}. In RQT, the rotation facilitates the alignment of quantized vortices along the rotation axis within a turbulent flow. A few studies on rotating thermal counter flow in superfluid $^4{\rm He}$ have investigated the competition between the established order due to the rotation and the disorder due to the turbulence \cite{Swanson1983,Tsubota2003,Tsubota2004}. Recent investigations into the decaying RQT in atomic BEC have revealed the anisotropic dissipation mechanisms \cite{Estrada2022} and a $k^{-1}$ scaling of the incompressible energy spectrum which is gentler than the $k^{-2}$ scaling in classical rapidly rotating turbulence \cite{Estrada2022,Estrada2022_2}. Thus, the RQT displays distinct characteristics compared to the classical case, which is attributed to the unique features of quantum systems, such as quantized vortices and Kelvin wave cascades.

However, there is currently a lack of understanding regarding the emergence of isotropy in RQT. In particular, the mechanism by which a direct turbulent cascade produce isotropy in the presence of rotation causing anisotropy is an important problem. Therefore, we consider a forced decaying RQT confined within a harmonic potential and conduct numerical investigations on the emergence of isotropy during its development using the GP model.

\section{\label{sec:model} Theoretical model}

\subsection{\label{sec:numerical} Gross-Pitaevskii model}
Our theoretical model is governed by the GP equation in a rotating frame with an angular velocity denoted by $\boldsymbol{\Omega}=(0,0,\Omega_z)$. The dimensionless equation is defined as follows:
\begin{eqnarray}
i\frac{\partial \psi(\boldsymbol{r},t)}{\partial t} &=&[-\nabla^2+V_{\rm har}(\boldsymbol{r})+V_{\rm for}(\boldsymbol{r},t)+g|\psi(\boldsymbol{r},t)|^2\notag \\ 
&&-\Omega_z\hat{L}_z(\boldsymbol{r})]\psi(\boldsymbol{r},t),
\label{eq:GP}
\end{eqnarray}
where $\psi(\boldsymbol{r},t)$ represents the macroscopic wavefunction of the condensate, and $g$ denotes the coupling constant. Here, the length scale and time are normalized by the characteristic scale $\tilde{a}_0=\sqrt{\hbar/(2\tilde{m}\tilde{\omega})}$ and time $1/\tilde{\omega}$, where $\tilde{m}$ is the atomic mass and $\tilde{\omega}$ represents the trapping frequency of the harmonic potential. The rotational term is expressed by the angular momentum operator $\hat{L}_z(\boldsymbol{r})=-i(x\partial_y-y\partial_x)$ and the angular frequency $\Omega_z$. The trapping potential is characterized as a weakly elliptical harmonic potential, denoted by 
\begin{equation}
V_{\rm har}(\boldsymbol{r})=\frac{1}{4}[(1+\delta)x^2+(1-\delta)y^2+z^2],
\label{eq:har}
\end{equation}
where $\delta$ denotes an elliptical deformation parameter. To induce turbulence, we introduced a random forcing potential
\begin{equation}
V_{\rm for}(\boldsymbol{r},t)=A[W(\boldsymbol{r},t)]
\label{eq:for}
\end{equation}
into the system. Here, $A$ is the normalization factor of the amplitude of random distribution $W(\boldsymbol{r},t)$. The Fourier transform of $W(\boldsymbol{r},t)$ is defined as $f_k\exp\{i\sigma(\boldsymbol{k},t)\}$, where $\sigma(\boldsymbol{k},t)$ represents uniform random numbers within the range of $[0,2\pi)$, and $f_k$ is unity within $k_f-\delta k/2\le k\le k_f+\delta k/2$ ($\delta k$ is the width of the energy injection) and is zero otherwise \cite{Proment2012,Shukla2022}. In this study, the forcing wavenumber $k_f$ is set to $\pi/R_{\rm TF}$, where $R_{\rm TF}$ denotes the Thomas-Fermi radius of the condensate trapped by $V_{\rm har}(\boldsymbol{r})$. We place random numbers of $V_{\rm for}(\boldsymbol{r},t)$ obtained by the above manner at a time interval $\Delta T=1$ and connect smoothly them at each spatial point by using the temporal interpolation with the natural cubic spline. Therefore, this potential injects energy at a low wavenumber $k_f$, thereby initiating a direct cascade, as elaborated in the subsequent discussions.

In this model, the total energy is expressed as $E(t)=E_{\rm kin}(t)+E_{\rm pot}(t)+E_{\rm non}(t)+E_{\rm rot}(t)$. Here, each energy component is defined as
\begin{eqnarray}
&&E_{\rm kin}(t)=\int |\nabla\psi|^2{\rm d}\boldsymbol{r},\ E_{\rm pot}(t)=\int (V_{\rm har}+V_{\rm for})|\psi|^2{\rm d}\boldsymbol{r},\notag \\
&&E_{\rm non}(t)=\int \frac{g}{2}|\psi|^4{\rm d}\boldsymbol{r},\ E_{\rm rot}(t)=-\Omega_z\int \psi^* \hat{L}_z\psi{\rm d}\boldsymbol{r},
\label{eq:kin}
\end{eqnarray}
where $E_{\rm kin}(t)$ is the kinetic energy, $E_{\rm pot}(t)$ the potential energy, $E_{\rm non}(t)$ the nonlinear energy, and $E_{\rm rot}(t)$ is the rotational energy. It is widely acknowledged that the kinetic energy can be decomposed through the Madelung transformation $\psi(\boldsymbol{r},t)=\sqrt{n(\boldsymbol{r},t)}\exp\{i\theta(\boldsymbol{r},t)\}$ (where $n(\boldsymbol{r},t)$ is the density and $\theta(\boldsymbol{r},t)$ is the phase) \cite{nore1997,nore1997fluids}, yielding $E_{\rm kin}(t)=E_q(t)+E^i(t)+E^c(t)$, where the components are defined as
\begin{equation}
E_{\rm q}(t)=\int \left(\nabla\sqrt{n(\boldsymbol{r},t)}\right)^2 {\rm d}\boldsymbol{r}
\end{equation}
and
\begin{equation}
E^{i,c}(t)=\int \left[\sqrt{n(\boldsymbol{r},t)}\boldsymbol{v}^{i,c}(\boldsymbol{r},t)\right]^2 {\rm d}\boldsymbol{r}.
\label{eq:qic}
\end{equation}
Here, $\boldsymbol{v}(\boldsymbol{r},t)=\nabla\theta(\boldsymbol{r},t)$ denotes the velocity, $[\cdots]^i$ represents the incompressible part with $\nabla \cdot [\cdots]=0$, and $[\cdots]^c$ denotes the compressible part with $\nabla \times [\cdots]=0$.

\subsection{\label{sec:energy} Numerical method}
To investigate the dynamics of forced decaying RQT, we numerically solved the equation
\begin{eqnarray}
[i-\gamma(\boldsymbol{k})]\frac{\partial }{\partial t}\Psi(\boldsymbol{k},t)=k^2\Psi(\boldsymbol{k},t)+H(\boldsymbol{k},t),
\label{eq:GP_num}
\end{eqnarray}
where $\Psi(\boldsymbol{k},t)$ and $H(\boldsymbol{k},t)$ denote the Fourier transforms of $\psi(\boldsymbol{r},t)$ and $[V_{\rm har}(\boldsymbol{r})+V_{\rm for}(\boldsymbol{r},t)+g|\psi(\boldsymbol{r},t)|^2-\Omega_z\hat{L}_z(\boldsymbol{r})]\psi(\boldsymbol{r},t)$, respectively. The damping term $\gamma(\boldsymbol{k})$ emulates the influence of the thermal excitation, as demonstrated in a previous study \cite{Kobayashi2006}. Based on the numerical findings at $0.001T_c$ (where $T_c$ denotes the critical temperature of an ideal Bose gas) \cite{Kobayashi2006}, we adopted the wavenumber dependence as $\gamma(k)=\gamma_0\theta(k-k_\xi)$. Here, $\theta(k)$ represents the step function and $k_\xi=2\pi \tilde{a_0}/\tilde{\xi}$ is the wavenumber corresponding to the healing length $\tilde{\xi}$. This damping term effectively suppresses the compressible sound waves at wavenumbers greater than $k_\xi$ \cite{Kobayashi2005,Kobayashi2005jps}. Therefore, in this system, neither the total number of particles $N$ nor the energy is conserved owing to the presence of external forcing and dissipation.

\begin{figure}[b]
\includegraphics[width=8.6cm]{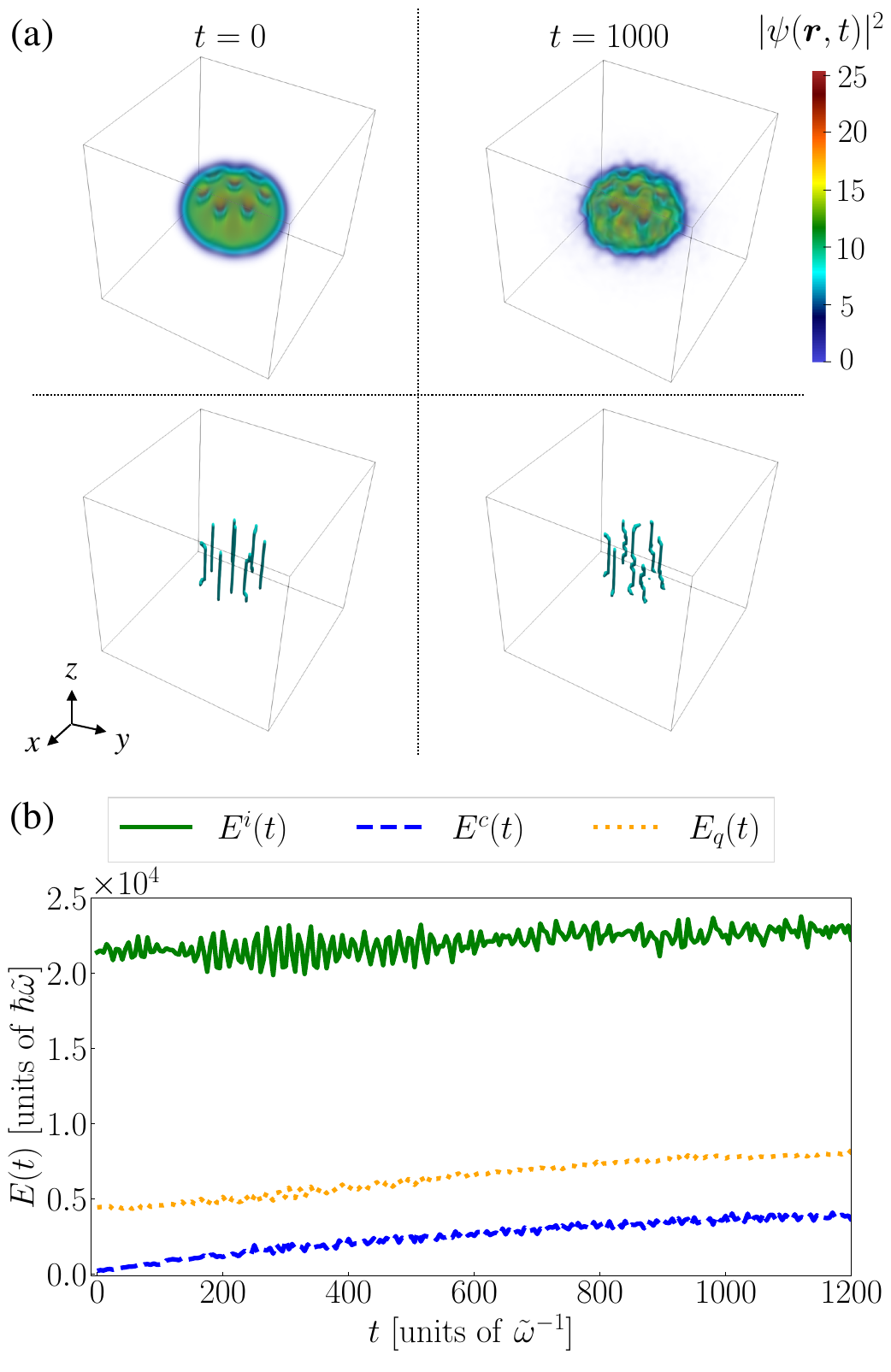}
\caption{\label{fig:ini_evo} Emergence of turbulent state. (a) Density distributions (upper panels) and configurations of quantized vortices inside the Thomas-Fermi radius $R_{\rm TF}$ (lower panels) at $t=0$ and $t=1000$. The color bar shows $|\psi(\boldsymbol{r},t)|^2$ in the unit of $\tilde{a}_0^{-3}$. The vortices are visualized by connecting the points where the phase rotates by $2\pi$. (b) Time evolution of the quantum $E_q(t)$, incompressible $E^i(t)$, and compressible $E^c(t)$ energies.}
\end{figure}

Our computational grid has a size of $V_{\rm num}=(L_{\rm num})^3=25^3$, with $N_{\rm grid}=128^3$ grid points. In our simulation, we set the numerical parameters to $g=0.28$ and $N=1.25\times 10^4$ based on experimental findings \cite{madison2000,Pethick2001}. Numerical simulations are conducted using the pseudo-spectral method employing a fourth-order Runge-Kutta time evolution with a time resolution of $10^{-3}$. The initial state corresponds to the equilibrium vortex-lattice state in a rotating BEC at $\Omega_z=0.70$ trapped by $V_{\rm har}(\boldsymbol{r})$ with a small elliptical deformation $\delta=0.025$. Using this initial state, we solve Eq.~(\ref{eq:GP_num}) with a nonzero forcing amplitude $A$ and $\gamma_0=0.06$ to investigate the turbulence evolution.

For convenience, in subsequent discussions, we introduce a spherically averaged spectrum in a discrete wavenumber space defined by
\begin{eqnarray}
g(k,t)=\frac{1}{u(k)}\sum_{\boldsymbol{k}'\in \Omega_k} G(\boldsymbol{k}',t),
\label{eq:ene_k}
\end{eqnarray}
where $G(\boldsymbol{k},t)$ is an arbitrary real function, $\Omega_k=\{\boldsymbol{k}'\,|\,k-\delta k/2\le |\boldsymbol{k}'| < k+\delta k/2\}$, $\delta k=2\pi/L_{\rm num}$, and $u(k)=\sum_{\boldsymbol{k}'\in \Omega_k} 1$ is the number of grid points in $\Omega_k$. Using the definition above, we represent the spherically averaged spectrum of a three-dimensional real function as the lowercase character of that function.

\section{\label{sec:D_turbulence} Numerical results}

\begin{figure}[b]
\includegraphics[width=8.6cm]{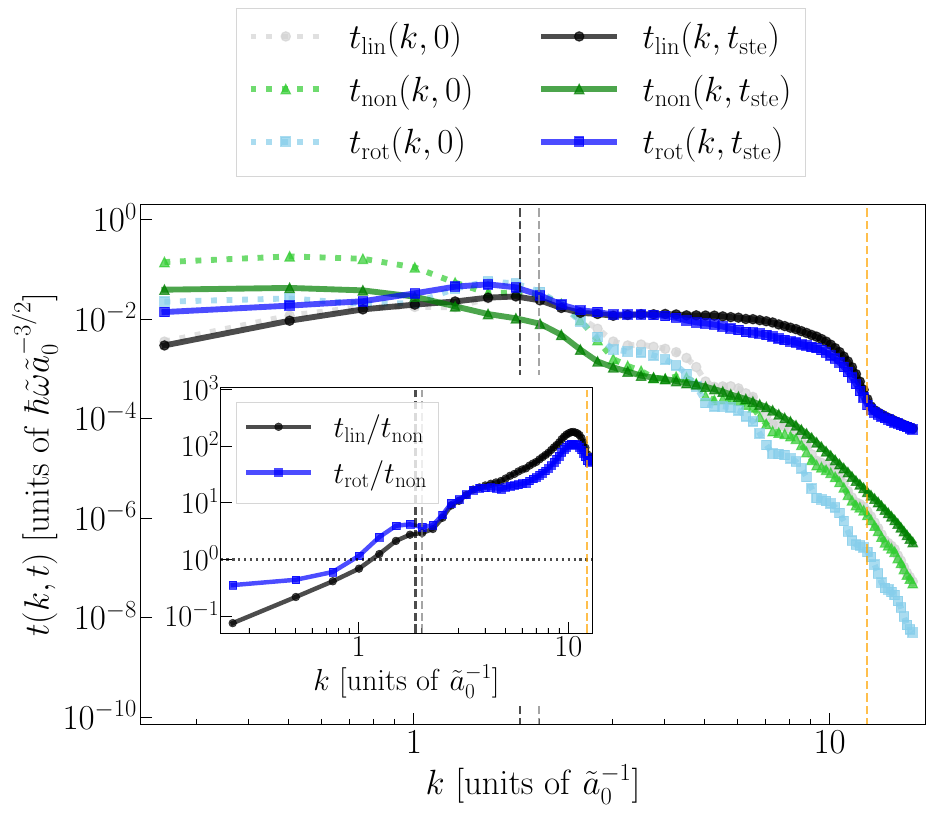}
\caption{\label{fig:GP_spe} Spectra of the linear $t_{\rm lin}(k,t)$ (black), nonlinear $t_{\rm non}(k,t)$ (green), and rotational $t_{\rm rot}(k,t)$ (blue) terms in the GP equation. The light colored lines show the spectra of the initial state at $t=0$ and dark colored ones show them of the turbulent state at $t=t_{\rm ste}$. The vertical black, gray and orange dotted lines correspond to $k_{\rm int }(0)=1.87$, $k_{\rm int }(t_{\rm ste})=2.00$, and $k_{\xi}=12.3$, respectively. Inset; the ratio of $t_{\rm lin}(k,t)$ and $t_{\rm rot}(k,t)$ to $t_{\rm non}(k,t)$. }
\end{figure}

\begin{figure*}[t]
\includegraphics[width=14cm]{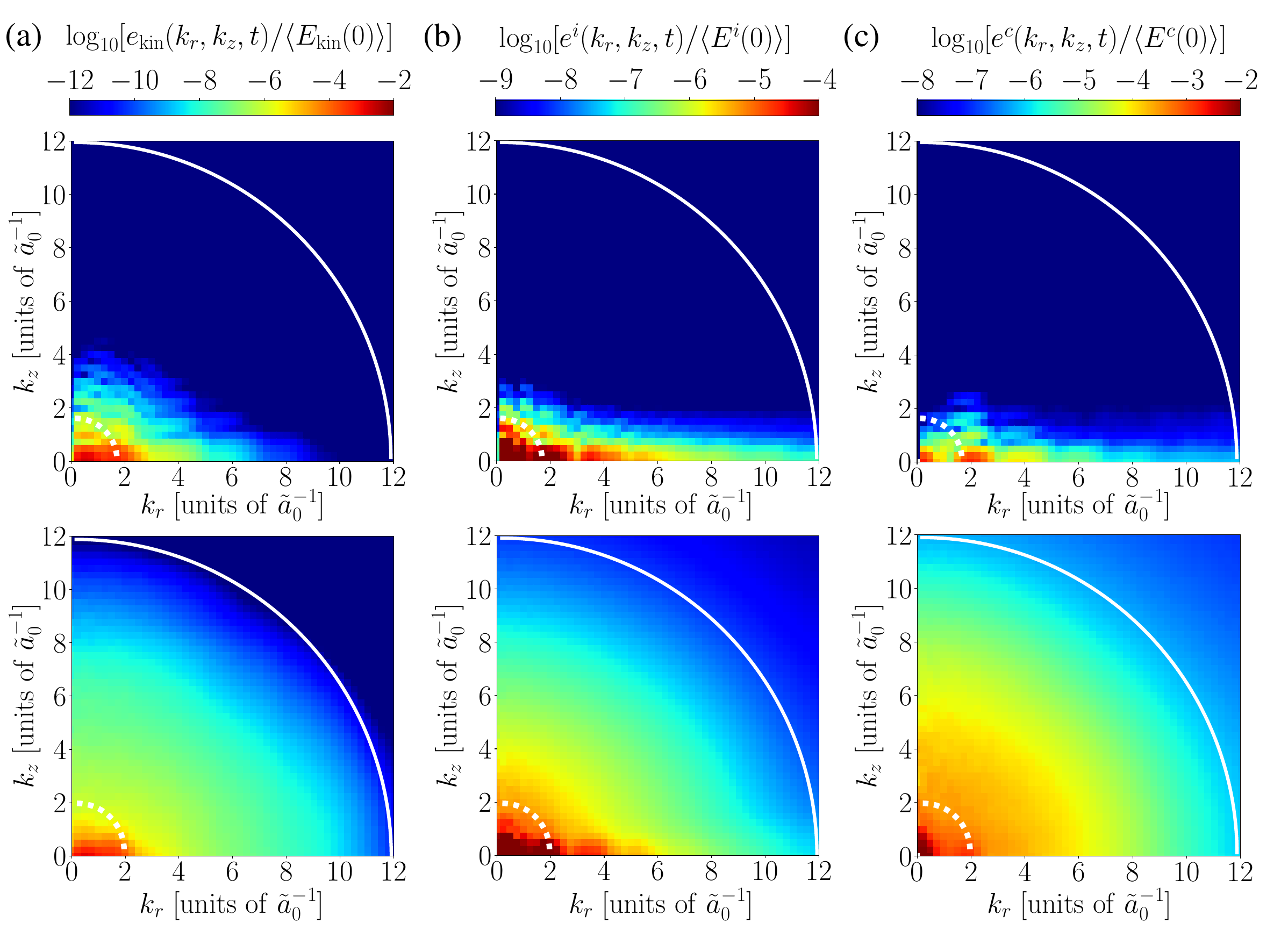}
\caption{\label{fig:avedis} Averaged energy distributions in the $k_r \mathchar`- k_z$ plane at the initial $t=0$ (upper panels) and turbulent state $t=t_{\rm ste}$ (lower panels). (a)--(c) indicate the normalized distributions of kinetic $e_{\rm kin}(k_r,k_z,t)$, incompressible $e^i(k_r,k_z,t)$, and compressible energies $e^c(k_r,k_z,t)$, respectively. Here, the normalization factor $ \langle \rangle$ denotes spatial average in the system and the distributions are plotted on a logarithmic scale. The white dotted (resp. solid) lines corresponds to the radius of $k_{\rm int}(t)$ (resp. $k_\xi$).}
\end{figure*}

We perform simulations at a finite forcing amplitude $A=0.25$, as shown in Fig.~\ref{fig:ini_evo}. The upper and lower panels in Fig.~\ref{fig:ini_evo}(a) illustrate the density distributions $|\psi(\boldsymbol{r},t)|^2$ and the spatial distributions of quantized vortices at $t=0$ and $t=1000$, respectively. The initial lattice state, including eight quantized vortices, transitions to a disordered turbulent state at $t=1000$ by the random forcing. Consequently, the number of vortices decreases from eight to seven, while the vortex line density increases from $0.088$ at $t=0$ to $0.101$ at $t=1000$. This comes from the distortion of the quantized vortices caused by the turbulent motion, and the helical structure of the coherent vortices becomes more clear than the classical rotating turbulence \cite{Bartello1994}. To identify the development of turbulence in terms of the kinetic energy, we present the time evolution of the quantum $E_q(t)$, incompressible $E^i(t)$, and compressible $E^c(t)$ energies in Fig.~\ref{fig:ini_evo}(b). During the initial stage, $E_q(t)$ and $E^c(t)$ increase by the excitation of intricate waves due to the random forcing, while $E^i(t)$ oscillates around its initial value. In the later phase, all the energy components become statistically steady at approximately $t_{\rm ste} \equiv 1000$. For reference, the particle loss at $t=t_{\rm ste}$ is only $0.2\%$ of the initial number of particles.

We next discuss the magnitudes of the linear, nonlinear, and rotational terms in the GP equations. Figure~\ref{fig:GP_spe} shows spherically averaged spectra $t_{\rm lin}(k,t)$, $t_{\rm non}(k,t)$, and $t_{\rm rot}(k,t)$ of $|T_{\rm lin}(\boldsymbol{k},t)|$, $|T_{\rm non}(\boldsymbol{k},t)|$, and $|T_{\rm rot}(\boldsymbol{k},t)|$ in the initial (at $t=0$) and turbulent (at $t=1000$) states. Here, $T_{\rm lin}(\boldsymbol{k},t)$, $T_{\rm non}(\boldsymbol{k},t)$, and $T_{\rm rot}(\boldsymbol{k},t)$ are the Fourier transforms of the kinetic ($=\nabla^2 \psi$), interaction ($=g|\psi|^2\psi$), and rotational ($=-\Omega_z\hat{L}_z\psi$) terms respectively. Reflecting the spatial distribution of vortices, $t_{\rm rot}(k,t)$ exhibits a higher magnitude around $k_{\rm int}(t)$ compared with other regions of wavenumbers, where $k_{\rm int}(t)$ is the wavenumber corresponding to the mean intervortex distance. In the turbulent state, $t_{\rm lin}(k,t)$ and $t_{\rm rot}(k,t)$ surpass $t_{\rm non}(k,t)$ within the intermediate wavenumber range $3 \lesssim k < k_\xi$, as shown in the inset. Therefore, in this regime, the interaction between waves is relatively weak, and the dynamics exhibit a weak rotating wave turbulent behavior.

 \begin{figure*}[t]
\includegraphics[width=\linewidth]{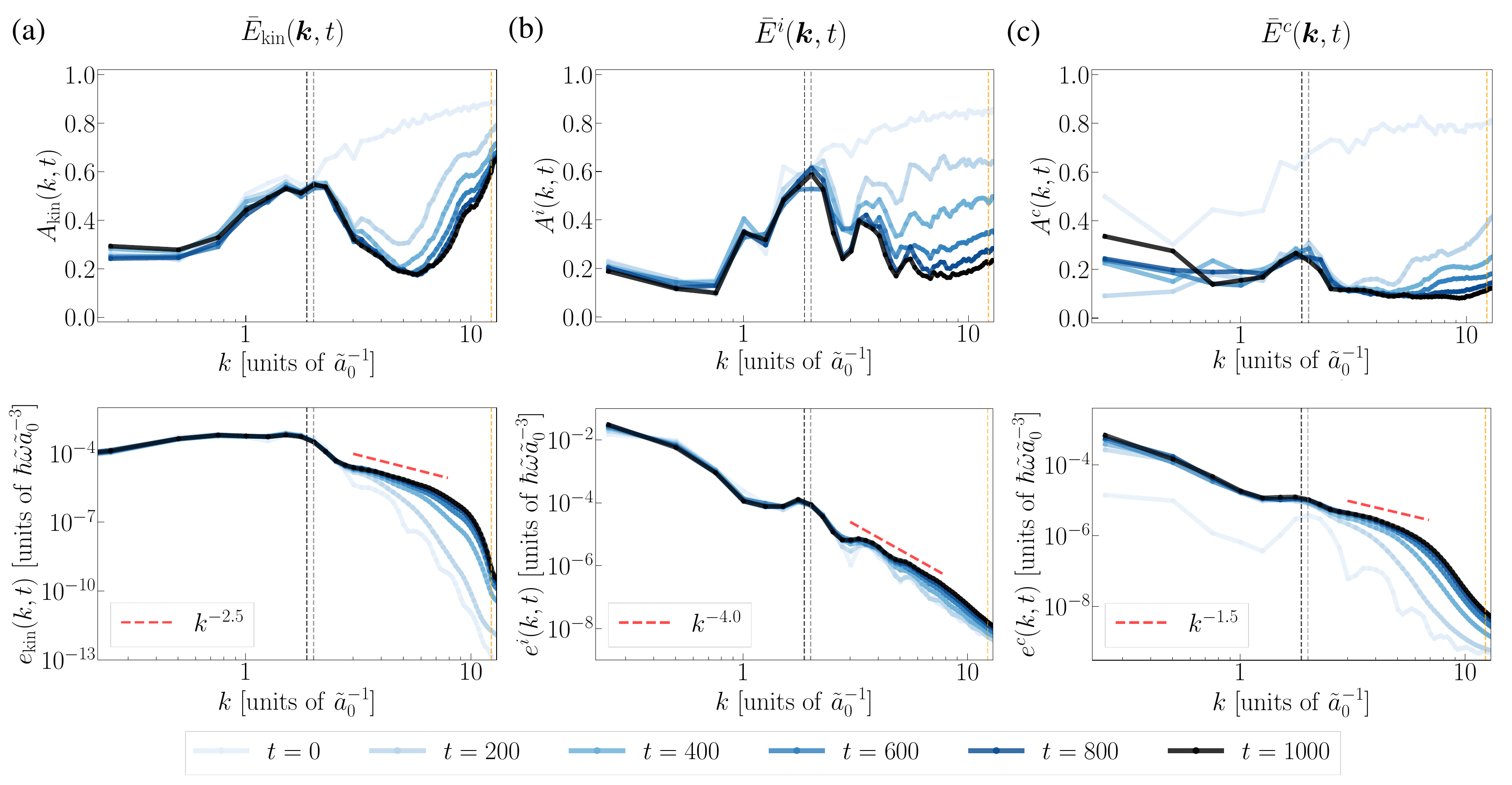}
\caption{\label{fig:anispe} Time evolution of anisotropy (upper panels) and spherically averaged energy spectrum (lower panels). (a)--(c) show the kinetic, incompressible, and compressible energies, respectively. Each spherically averaged energy spectrum is defined as follows: $e_{\rm kin}(k,t)=\sum_{\boldsymbol{k}'\in \Omega_k} \bar{E}_{\rm kin}(\boldsymbol{k}',t)/u(k)$, $e^i(k,t)=\sum_{\boldsymbol{k}'\in \Omega_k} \bar{E}^i(\boldsymbol{k}',t)/u(k)$, and $e^c(k,t)=\sum_{\boldsymbol{k}'\in \Omega_k} \bar{E}^c(\boldsymbol{k}',t)/u(k)$. Each power law within the range of $3\le k \le 8$ is indicated as red dashed lines. The $k_{\rm int}(0)$, $k_{\rm int}(t_{\rm ste})$, and $k_{\xi}$ are shown in black, gray, and orange, respectively.}
\end{figure*}

In classical rotating turbulence, it is well known that dynamics exhibit anisotropic characteristics within the wavenumber region where the Coriolis force is dominant. Drawing an analogy, we assume that the energy distribution of RQT is anisotropic in the region where the rotational effect predominates over the nonlinear effect. To ascertain the anisotropy, we visualize the $k_r \mathchar`- k_z$ dependence of each energy distribution at $t=0$ and $t=t_{\rm ste}$ in Fig.~\ref{fig:avedis}, where $k_r=(k_x^2+k_y^2)^{1/2}$. The averaged kinetic energy distribution around the $k_z$ axis is defined as
 \begin{equation}
e_{\rm kin}(k_r,k_z,t)=\frac{1}{w(k_r)}\sum_{\boldsymbol{k}' \in S_k} \bar{E}_{\rm kin}(\boldsymbol{k}',t).
\label{eq:ene_krz}
\end{equation}
Here, $S_k=\{\boldsymbol{k}'\,|\,k_r-\delta k/2\le k_r' < k_r+\delta k/2\}$, $w(k_r)=\sum_{\boldsymbol{k}' \in S_k} 1$, $\overline{[\cdots]}$ denotes the time-averaged distribution over the time scale $4\Delta T$ to reduce the fluctuation with the energy injection, and other energy distributions are defined in a similar manner. The upper panels show the averaged energy distributions at $t=0$. These energy distributions are anisotropic, with the energy predominantly concentrated around $k_{\rm int}(0)$ on the $k_r$ axis, reflecting the initial vortex lattice. The lower panels display the distributions at $t=t_{\rm ste}$. The initial anisotropic distribution transitions towards greater isotropy with an increase in the high wavenumber components. However, $e_{\rm kin}(k_r,k_z,t)$ and $e^i(k_r,k_z,t)$ retain their anisotropic characteristics at approximately $k_{\rm int}(t_{\rm ste})$. This anisotropy originated from a quasi-two-dimensional flow characterized by quantized vortices. By contrast, $e^c(k_r,k_z,t)$ shows an isotropic distribution at $k> k_{\rm int}(t_{\rm ste})$, indicating that the compressible wave motion at small scales is slightly affected by anisotropy owing to rotation. Additionally, we observe a decay in the energies within the dissipative region $k\ge k_\xi$ outside the white solid line.

We now turn to the study of the temporal emergence of isotropy by employing a quantitative measure of anisotropy $A(k,t)$, as introduced in our previous study \cite{Sano2022}. This measure quantitatively evaluates the anisotropy of distribution on a spherical surface of radius $k$, yielding values ranging from zero to unity. If the distribution is perfectly isotropic at $k$, $A(k,t)$ vanishes. 

\begin{figure}[b]
\includegraphics[width=8.6cm]{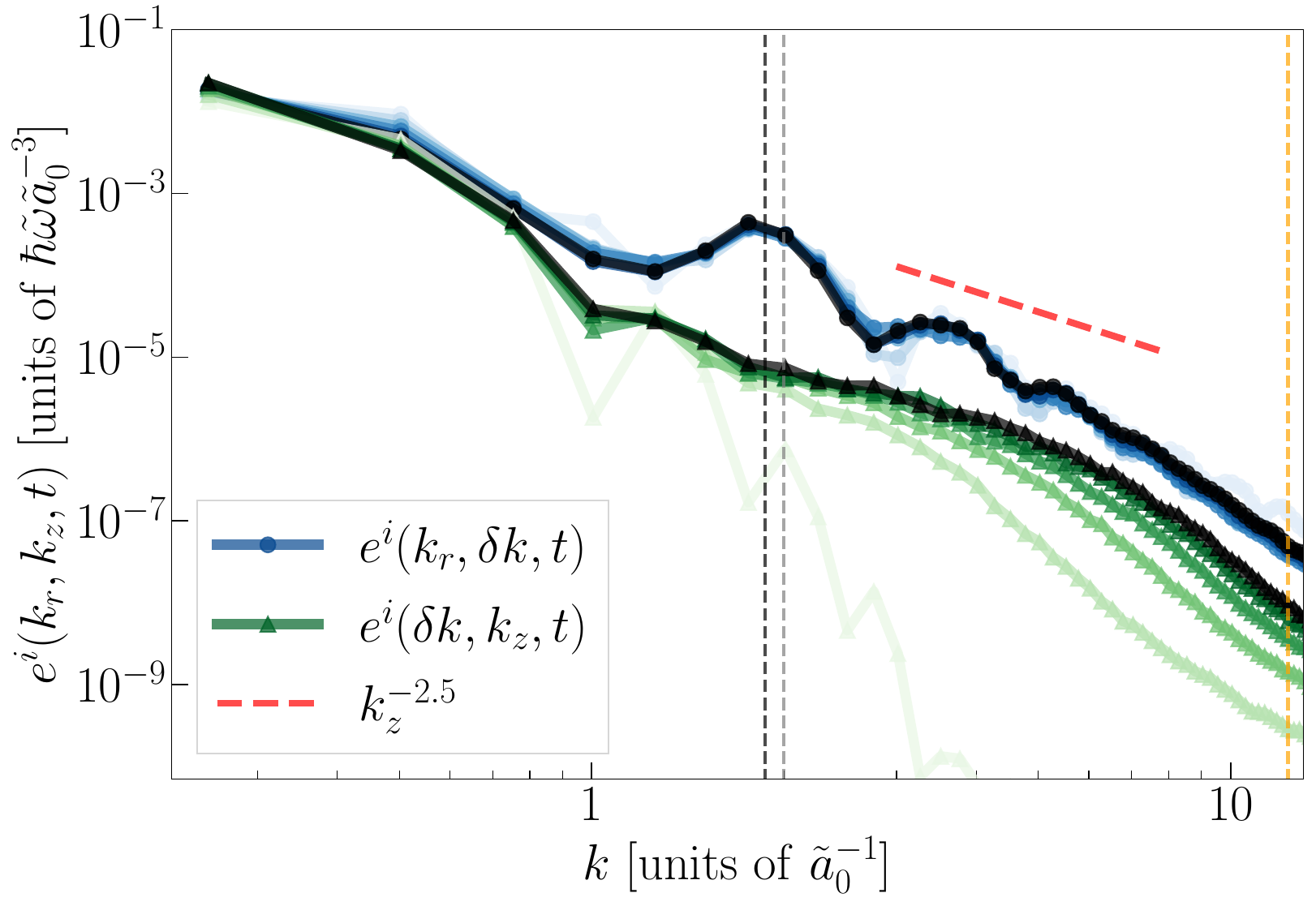}
\caption{\label{fig:in_ani} Time evolution of the averaged incompressible kinetic energy spectrum $e^i(k_r,k_z,t)$ at $k_r=\delta k$ (green line) and $k_z=\delta k$ (blue line). The change of color corresponds to the time evolution shown in Fig.~\ref{fig:anispe}. Power law and characteristic wavenumbers are indicated as references. }
\end{figure}

Figure~\ref{fig:anispe} illustrates the temporal evolution of the anisotropies $A_{\rm kin}(k,t)$,  $A^i(k,t)$, and $A^c(k,t)$ of $\bar{E}_{\rm kin}(\boldsymbol{k},t)$, $\bar{E}^i(\boldsymbol{k},t)$, and $\bar{E}^c(\boldsymbol{k},t)$ in the upper panels. The lower panels show the spherically averaged spectra $e_{\rm kin}(k,t)$, $e^i(k,t)$, and $e^c(k,t)$ for these energy distributions. The pronounced influence of the energy injection around $k_f=0.62$ becomes apparent in $e^c(k,t)$ whereas $e_{\rm kin}(k,t)$ and $e^i(k,t)$ do not change significantly around $k_f$. The injected energy propagates toward higher wavenumbers owing to nonlinear interactions. The direct turbulent cascade moderates the initial anisotropic peak structure of each energy component, resulting in a decrease in their anisotropies at $k>k_{\rm int}(t)$. Consequently, $e^c(k,t)$ exhibits an isotropic power-law behavior characterized by $k^{-1.5}$. Regarding $e_{\rm kin}(k,t)$, it demonstrates $k^{-2.5}$ dependence in $3\le k \le 8$; however, $A_{\rm kin}(k,t)$ displays high values around $k_{\rm int}(t_{\rm ste})$ and $k_{\xi}$. Thus, the scaling behavior depends on the wave vector direction. The power exponent of $k^{-4}$ in $e^i(k,t)$ is consistent with the $k^{-2}$ scaling observed in the spherically integrated spectrum of classical rotating turbulence in an incompressible flow \cite{Zhou1995,Canuto1997,Yeung1998,Baroud2002}. However, $A^i(k,t)$ exhibits significant magnitudes around wavenumbers which are integer multiples of $k_{\rm int} (t_{\rm ste})$. This anisotropy reflects the concentration of $e^i(k_r,k_z,t)$ on the $k_r$ axis caused by a weakly disrupted lattice structure of vortices, as shown in Fig.~\ref{fig:avedis}(c). Therefore, the scaling of $e^i(k,t)$ also exhibits directional dependence.

To validate the anisotropy of the incompressible kinetic energy distribution, we present the averaged energy spectrum $e^i(k_r,k_z,t)$ near the $k_r$ and $k_z $ axes in Fig.~\ref{fig:in_ani}. In the $k_z$ direction, a significant increase can be observed, whereas there is slight change in the $k_r$ direction, which indicates the strong anisotropy of $e^i(k_r,k_z,t)$. We attribute this anisotropy to the larger energy flux along the $k_z$ axis than in the $k_r$ axis. The anisotropic turbulent cascade forms a $k_z^{-2.5}$ scaling near the $k_z$ axis, and this power exponent is slightly steeper than the obtained spectrum in rapidly rotating wave turbulence \cite{Bellet2006}. On the other hand, a power-law behavior does not emerges along the $k_r$ axis, and $e^i(k_r,\delta k,t)$ shows the characteristic dependence on $k_r$, reflecting the arrangement of vortices.

\section{\label{sec:conclusion} Conclusion}
We focused on the effect of rotation on the anisotropy of quantum turbulence, and investigated the development of forced decaying RQT trapped by a weakly elliptical harmonic potential using the GP model. During the development of RQT, the alignment of quantized vortices is weakly disrupted by the turbulent flow, and the anisotropy of kinetic energy distribution is softened. In the wavenumber region, where the dominance of the rotation term in the GP equation surpasses that of the nonlinear term, we found the anisotropic development of the incompressible kinetic energy distribution and the obtained average spectrum displays $k_z^{-2.5}$ near the $k_z$ axis. By contrast, the compressible kinetic energy distribution exhibits an isotropic power law of $k^{-1.5}$. 

Compared to the discovery of the emergent isotropy in non-rotating quantum turbulence \cite{Galka2022,Sano2022}, this study clearly demonstrated the manifestation of an anisotropic turbulent cascade due to the rotation. In the future, we aim to systematically explore the anisotropic turbulent cascade in RQT by varying the angular frequency of the rotation and the magnitude of the external forcing.

\section{\label{sec:acknowledgments} Acknowledgments}
Y.~S. acknowledges the support from JST and the establishment of university fellowships towards the creation of science technology innovation (Grant No. JPMJFS2138). M.~T. acknowledges support from JSPS KAKENHI (Grant No. JP23K03305).

\nocite{*}


\end{document}